\documentclass[twocolumn,english,aps,prl,superscriptaddress]{revtex4-1}
\usepackage[T1]{fontenc}
\usepackage[latin9]{inputenc}
\usepackage{color}
\usepackage{refstyle}
\usepackage{amsmath}
\usepackage{amssymb}
\usepackage{graphicx}
\usepackage{subscript}

\makeatletter

\AtBeginDocument{\providecommand\figref[1]{\ref{fig:#1}}}
\RS@ifundefined{subsecref}
  {\newref{subsec}{name = \RSsectxt}}
  {}
\RS@ifundefined{thmref}
  {\def\RSthmtxt{theorem~}\newref{thm}{name = \RSthmtxt}}
  {}
\RS@ifundefined{lemref}
  {\def\RSlemtxt{lemma~}\newref{lem}{name = \RSlemtxt}}
  {}

\newcommand{\diagram}[1]{\vcenter{\hbox{\includegraphics[scale=0.5]{#1}}}}
\renewcommand{\fnum@figure}{FIG.~\thefigure}
\newref{fig}{name = {Fig.~}}
\usepackage{hyperref}
\hypersetup{colorlinks=true,linkcolor=blue,citecolor=blue,urlcolor=blue}
\usepackage{bibunits}
\usepackage{titlesec}
\titleformat*{\section}{\large\bfseries}
\usepackage{chngcntr}
\counterwithin*{equation}{section}
\usepackage{bibunits}

\makeatother

\usepackage{babel}
\begin{document}
\begin{bibunit}[apsrev4-1]

\title{Dichotomy between in-plane magnetic susceptibility and resistivity
anisotropies in extremely strained BaFe\textsubscript{2}As\textsubscript{2} }

\author{Mingquan He}
\email{mingquan.he@kit.edu}

\selectlanguage{english}%

\affiliation{\textsuperscript{}Institute for Solid State Physics, Karlsruhe Institute
of Technology, 76021 Karlsruhe, Germany}

\author{Liran Wang}

\affiliation{\textsuperscript{}Institute for Solid State Physics, Karlsruhe Institute
of Technology, 76021 Karlsruhe, Germany}

\author{Felix Ahn}

\affiliation{\textsuperscript{}Institut für Theoretische Physik III, Ruhr-Universität
Bochum, D-44801 Bochum, Germany}

\author{Frédéric Hardy}

\affiliation{\textsuperscript{}Institute for Solid State Physics, Karlsruhe Institute
of Technology, 76021 Karlsruhe, Germany}

\author{Thomas Wolf}

\affiliation{\textsuperscript{}Institute for Solid State Physics, Karlsruhe Institute
of Technology, 76021 Karlsruhe, Germany}

\author{Peter Adelmann}

\affiliation{\textsuperscript{}Institute for Solid State Physics, Karlsruhe Institute
of Technology, 76021 Karlsruhe, Germany}

\author{Jörg Schmalian}

\affiliation{\textsuperscript{}Institute for Solid State Physics, Karlsruhe Institute
of Technology, 76021 Karlsruhe, Germany}

\affiliation{Institute for Theory of Condensed Matter, Karlsruhe Institute of Technology,
76131 Karlsruhe, Germany}

\author{Ilya Eremin}

\affiliation{\textsuperscript{}Institut für Theoretische Physik III, Ruhr-Universität
Bochum, D-44801 Bochum, Germany}

\author{Christoph Meingast}
\email{christoph.meingast@kit.edu}

\selectlanguage{english}%

\affiliation{\textsuperscript{}Institute for Solid State Physics, Karlsruhe Institute
of Technology, 76021 Karlsruhe, Germany}

\date{10/18/16}
\begin{abstract}
The in-plane resistivity and uniform magnetic susceptibility anisotropies
of BaFe\textsubscript{2}As\textsubscript{2} are obtained with a
new method, in which a large symmetry-breaking uniaxial strain is
applied using a substrate with a very anisotropic thermal expansion.
The resistivity anisotropy and its corresponding elastoresistivity
exhibit very similar diverging behavior as those obtained from piezo-stack
experiments. This suggests that the resistivity anisotropy is more
a direct measure of magnetism than of nematicity, since the nematic
transition is no longer well-defined under a large strain. In strong
contrast to the large resistivity anisotropy above $T_{N}$, the anisotropy
of the in-plane magnetic susceptibility develops largely below $T_{N}$.
Using an itinerant model, we show that the observed anisotropy ($\chi_{b}>\chi_{a}$)
is determined by spin-orbit coupling and the orientation of the magnetic
moments in the antiferromagnetic phase, and that the anisotropy is
dominated by intra-orbital ($yz,yz$) contributions of the Umklapp
susceptibility. 
\end{abstract}
\maketitle
One striking similarity between iron-based superconductors(IBS) and
high $T_{c}$ cuprate superconductors is that superconductivity emerges
in close proximity to a magnetic instability \cite{Ishida2009,Johnston2010,Hirschfeld2011}.
Most iron pnictides have a stripe-type antiferromagnetic phase, in
which the Fe magnetic moments are parallel to the ordering wave vector
either $\mathbf{Q}_{1}=\left(\pi,0\right)$ or $\mathbf{Q}_{2}=\left(0,\pi\right)$,
which breaks the $C_{4}$ symmetry of the paramagnetic structure \cite{Kitagawa2008,Huang2008,Dai:2012aa,Dai2015}.
The magnetic transition at $T_{N}$ is accompanied, or sometimes even
preceded, by a small orthorhombic structural distortion at $T_{S}\geqslant T_{N}$,
which has raised the question of whether magnetism alone is driving
these transitions \cite{Fang2008,Nandi2010,Fernandes2012,Fernandes:2014aa},
or whether orbital degrees of freedom also need to be considered \cite{Kruger2009,Kontani2010,Kontani2011,Yamase2013}.
This issue is particularly pressing for FeSe, which has no long-range
magnetic order down to the lowest temperature at ambient pressure
but nevertheless exhibits a similar orthorhombic distortion as the
other Fe-based materials \cite{Hsu2008,McQueen2009,Bohmer2013}. This
non-magnetic and orthorhombic phase has been coined 'electronic nematic'
\cite{Fradkin2010,Fernandes2012}. Experimentally, the susceptibility
to form a nematic state has been probed by a variety of methods, including
elastic \cite{Bohmer2014,Bohmer:2015aa,Bohmer:aa}, resistivity anisotropy
using a piezo stack \cite{Chu10082012,Kuo2013,Kuo2014,Kuo2015arXiv},
Raman scattering \cite{Gallais2013,Gallais2016,Kretzschmar:2016aa},
thermopower \cite{Jiang2013}, NMR \cite{Fu2012,Tetsuya2015}, optical
condutiviy \cite{Nakajima2011,Dusza2011}. Interestingly, many optimally
doped Fe-based materials appear to be close to a putative nematic
quantum critical point \cite{Kuo2015arXiv}, and recent theoretical
works suggest that electronic nematic fluctuations may provide a boost
to superconductivity in various channels \cite{Lederer2015}. 

In this Letter we study the interplay between magnetism and nematicity
in the parent compound BaFe\textsubscript{2}As\textsubscript{2}
using a somewhat different approach. Rather than probing the nematic
susceptibility in the zero-strain limit, we suppress the nematic transition
by imposing a large symmetry breaking strain on the crystal and then
examine the response of both the resistivity anisotropy and the magnetic
susceptibility anisotropy. In strong contrast to
the large resistivity anisotropy above $T_{N}$, the anisotropy of
the in-plane magnetic susceptibility develops largely below $T_{N}$,
although both quantities are to first-order expected to be proportional
to the nematic order parameter in the spin-nematic scenario \cite{Fernandes2012,Fernandes:2014aa}.
Further, we show that the resistivity anisotropy exhibits a sharp
maximum at $T_{N}$ and not at $T_{S}$, as expected in the spin-nematic
picture \cite{Fernandes2012,Fernandes:2014aa}. Using an itinerant
model, we show that the observed anisotropy ($\chi_{b}>\chi_{a}$)
is determined by spin-orbit coupling and the orientation of the magnetic
moments in the antiferromagnetic phase, and that the anisotropy is
dominated by intra-orbital ($yz,yz$) contributions of the Umklapp
susceptibility. 

\begin{figure}
\begin{centering}
\includegraphics[scale=0.5]{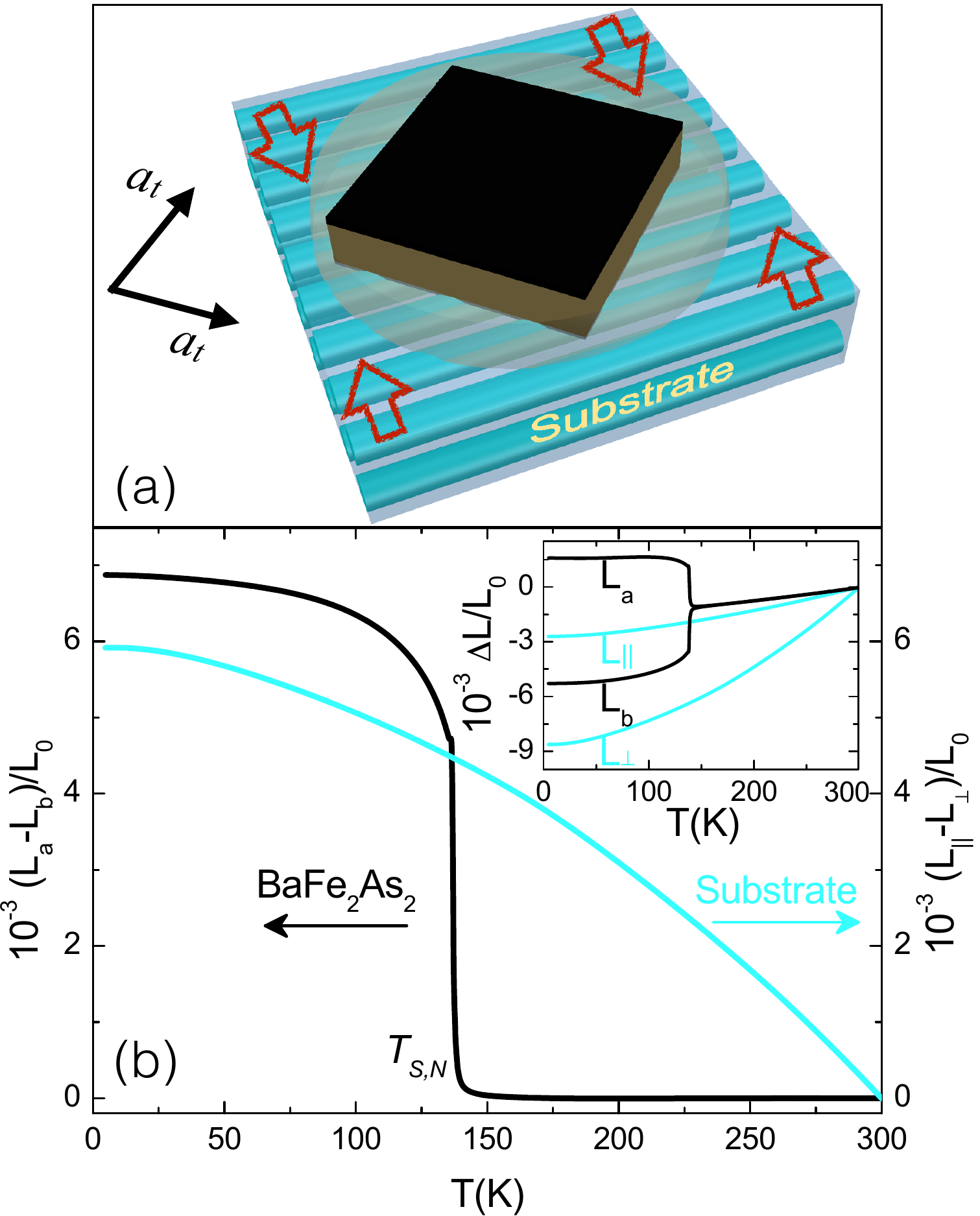}
\par\end{centering}
\caption{(a) Illustration of the uniaxial straining set-up. The crystal is
glued on top of a glass-fiber reinforced plastic substrate using epoxy
with the {[}110{]}$_{tet}$ direction parallel to fibers. Upon cooling,
the thermal-expansion anisotropy of the substrate applies a uniaxial
strain to the crystal. (b) Uniaxial strain of the substrate ($L_{\perp}$:
perpendicular to fibers, $L_{||}$: parallel to fibers) compared to
the in-plane orthorhombic distortion of a free standing BaFe\protect\textsubscript{2}As\protect\textsubscript{2 }crystal
($L_{a}$: longer orthorhombic axis, $L_{b}$: shorter orthorhombic
axis). The thermal expansion is shown in the inset. \label{fig:1}}
\end{figure}

Self-flux grown single crystals of BaFe\textsubscript{2}As\textsubscript{2},
with typical dimensions of 2 mm $\times$ 2 mm $\times$ 0.08 mm,
were glued onto a glass-fiber reinforced plastic (GFRP) substrate
using two-component epoxy(UHU Plus Endfest 300, 90 minutes) with the
crystal's tetragonal {[}110{]}$_{tet}$ direction orientated parallel
to the fibers (see Fig. \hyperref[fig:1]{1(a)} ). In order to determine
the uniaxial strain applied to the sample, the thermal expansion of
the GFRP substrate material was characterized by a home-built high
resolution capacitance dilatometer \cite{Meingast1990}. Electrical
contacts, with typical resistances of around 2$\varOmega$, were made
using silver paste, and the sample resistance along two perpendicular
directions was measured simultaneously on the same sample by a four-terminal
method. Magnetization measurements both parallel and perpendicular
to the fiber orientation of the substrate were carried out in a Physical
Property Measurement System (PPMS) using the Vibrating Sample Magnetometer
(VSM) unit from Quantum Design Inc. 

\begin{figure*}
\begin{centering}
\includegraphics[scale=0.5]{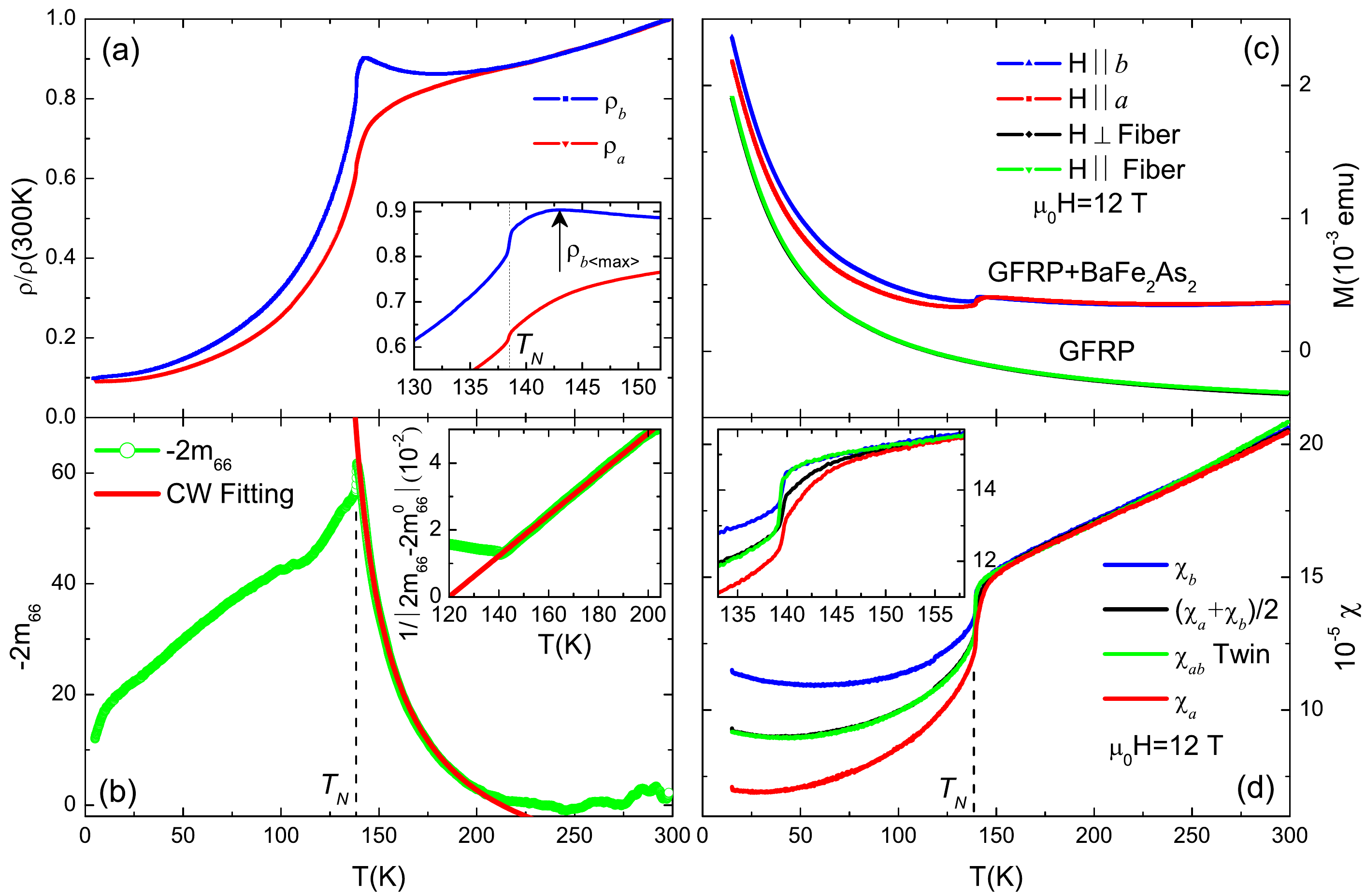}
\par\end{centering}
\caption{Temperature dependence of (a) the in-plane resistivity along a and
b directions, (b) the elastoresistivity tensor $2m_{66}$, (c) raw
magnetization data of GFRP alone and together with the BaFe\protect\textsubscript{2}As\protect\textsubscript{2}
crystal, and (d) anisotropic susceptibility obtained by subtracting
the GFRP background from the data shown in (c). The red solid line
in (b) is a Curie-Weiss fit ( $|2m_{66}|=a/(T-T_{0})+b$ with $T_{0}=120\pm1$
K ) and the inset shows the inverse plot. The insets in (a) and (d)
display magnified views near $T_{N}$. The arrow in the inset of (a)
indicates a maximum of $\rho{}_{b}$. \label{fig:2}}
\end{figure*}

Figure \hyperref[fig:1]{1(b)} shows that the difference of the thermal
expansion parallel and perpendicular to the fiber direction of the
substrate material is comparable in magnitude to the orthorhombic
distortion of a free standing BaFe\textsubscript{2}As\textsubscript{2}
crystal \cite{Bohmer:2015aa,Liran2016} near the transition temperature.
Thus, by glueing the BaFe\textsubscript{2}As\textsubscript{2} crystal
to this substrate at room temperature, a uniaxial symmetry-breaking
strain on the order of $\sim4\times10{}^{-3}$ can be expected at
140 K, which is roughly an order of magnitude larger than the strain
applied by the piezo-stack technique \cite{Chu10082012,Kuo2013,Kuo2014,Kuo2015arXiv}.
As will be shown in the following, our uniaxial straining technique
thus allows us to study the response under extreme conditions, and,
in particular, allows us to measure both the in-plane resistivity
and the uniform magnetic susceptibility anisotropies due to the small
size of the setup.

The measured in-plane resistivities of BaFe\textsubscript{2}As\textsubscript{2}
in the uniaxial-strain setup are shown in Figure \hyperref[fig:2]{2(a)}.
The resistivities $\rho{}_{b}$ and $\rho_{a}$ were measured on the
same sample and are normalized by the resistivity at 300 K in order
to eliminate geometrical uncertainties of the contacts. Our in-plane
resistivity anisotropy with $\rho{}_{b}>\rho_{a}$ is consistent with
the largest anisotropy $(\kappa=\rho_{b}/\rho_{a}-1)_{max}\sim40\%$
obtained by conventional detwinning methods \cite{Chu2010,Ishida2013,Tanatar2010,Ying2011,Blomberg:2013aa},
proving that the sample experiences a large uniaxial strain. A quite
high (for BaFe\textsubscript{2}As\textsubscript{2}) residual resistivity
ratio (RRR $\sim10$) is found, attesting for the high quality of
our crystals. The inset in Fig. \hyperref[fig:2]{2(a)} provides more
details near $T_{N}$. Both $\rho_{a}$ and $\rho{}_{b}$ exhibit
sharp drops at $T$=138.5 K, which we identify with the magnetic transition,
and $\rho{}_{b}$ has a maximum about 5 K above the magnetic transition.
The $m_{66}$ of the elastoresistivity tensor 
has proved very useful for studying the nematic susceptibility $\chi_{N}$
\cite{Chu10082012,Kuo2013,Kuo2014,Kuo2015arXiv}, can also be calculated
for our data since we know the applied anisotropic strain from the thermal expansion of the substrate (see \figref{1}). Here,
\begin{align}
2m_{66}(T) & =\frac{\rho_{b}(T)-\rho_{a}(T)}{\rho_{0}(T)(\varepsilon_{\perp}(T)-\varepsilon_{\parallel}(T))},\nonumber \\
 & \rho_{0}(T)=\frac{1}{2}\left[\rho_{b}(T)+\rho_{a}(T)\right].\label{eq:2}
\end{align}
We find (see Fig. \hyperref[fig:2]{2(b)}) that $|2m_{66}|$ exhibits
a very similar magnitude and divergent Curie-Weiss behavior as $T_{N}$
is approached from above as found in the elastoresistivity data obtained
using a piezo-stack, in which a much smaller strain is applied \cite{Chu10082012,Kuo2013,Kuo2014,Kuo2015arXiv}. 
This implies that the resistivity change $\Delta\rho/\rho_{0}(\varepsilon)$
varies approximately linearly with applied strain $\varepsilon$ up
to the large strains studied here. Similar to a ferromagnet in an
applied field, we no longer expect a real nematic phase transition
for the large strain applied here \cite{Fernandes2011}, and therefore
the observation of a sharp peak in the resistivity anisotropy is quite
surprising. Our results thus suggest that the resistivity anisotropy
is more directly related to the magnetic transition than to the nematic
fluctuations. We note that a similar conclusion can be deduced from
the data of Ref. \cite{Haoran2015}, in which the peak in the resistivity
anisotropy also occurs at $T_{N}$ in spite of the fairly large uniaxial
pressure applied.

Since the 'detwinning apparatus' in our case is reduced to a thin
substrate plate, our method is also feasible for investigating the
anisotropy of other quantities, e.g. the magnetization. Fig. \hyperref[fig:2]{2(c)}
displays the raw magnetization data at 12 Tesla of a BaFe\textsubscript{2}As\textsubscript{2}
crystal glued to the glass-fiber substrate in two different orientations
(blue and red lines), as well as the bare substrate in the same two
orientations (black and green lines). A clear sign of magnetization
anisotropy is already observable in the raw data below $T_{N}$, despite
of a considerable Curie-Weiss component in the magnetization of the
GFRP material, which needs to be subtracted. The calculated susceptibility
data after subtraction of the substrate background are shown in Fig.
\hyperref[fig:2]{2(d)} along with data of a free-standing crystal
in the twinned state. Well above $T_{N}$, the susceptibilities along
both directions are practically identical and decrease linearly with
temperature, as previously observed \cite{Wang2009} and also exists
in other Fe-based systems \cite{Zhang2009,Klingeler2010}. Below $T_{N}$,
the susceptibility along the longer axis $\chi_{a}$ becomes significantly
smaller than that of the shorter axis $\chi_{b}$. The difference
between $\chi_{a}$ and $\chi_{b}$ keeps increasing with decreasing
temperature and the anisotropic ratio $\eta=\chi_{b}/\chi_{a}-1$
reaches $\sim$60\% at 15 K. The average of $\chi_{a}$ and $\chi_{b}$
(black line in Fig. \hyperref[fig:2]{2(d)}) agrees excellently with
the twinned data $\chi_{t}$ within the whole temperature range, except
slightly above $T_{N}$ (see inset of Fig. \hyperref[fig:2]{2(d)}),
where the averaged data show a significant precursor to the transition
starting at about 150 K. We note that the observed sign, $\chi_{b}>\chi_{a}$,
explains the sign of the magnetic detwinning effect reported in Ref.
\cite{Chu2010field,Zapf2014}, however we observe no anisotropy at
$\sim$170 K, as claimed in torque magnetometry experiments on BaFe\textsubscript{2}As\textsubscript{2}
\cite{Kasahara:2012aa}. 

\begin{figure}[b]
\begin{centering}
\includegraphics[scale=0.28]{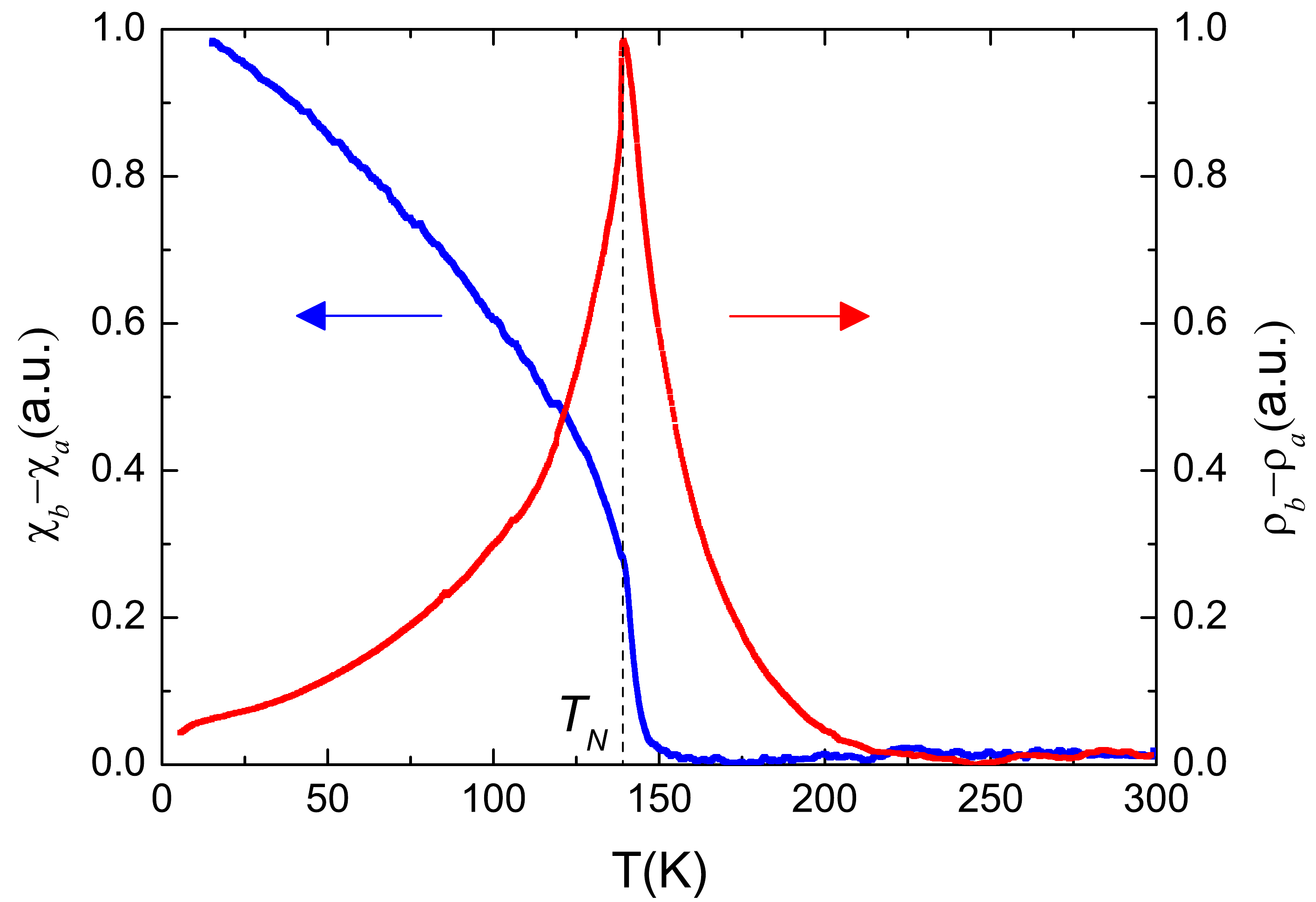}
\par\end{centering}
\caption{Temperature dependence of in-plane resistivity and susceptibility
anisotropies. Both curves are scaled for clarity.\label{fig:3}}
\end{figure}

\figref{3} highlights the surprisingly different behavior of the
susceptibility anisotropy, $\chi_{b}-\chi_{a}$, and the resistivity
anisotropy, $\rho{}_{b}-\rho_{a}$. Whereas $\rho{}_{b}-\rho_{a}$
is peaked close to and extends considerably above $T_{N}$, $\chi_{b}-\chi_{a}$
only starts to develop slightly above $T_{N}$ and then increases
to the lowest temperatures. Thus, the resistivity anisotropy and the
susceptibility anisotropy do not scale linearly with each other above
the transition, in contrary to the expectation of the spin-nematic
scenario \cite{Fernandes2012,Fernandes:2014aa}. Below
we show that this is due to the fact that the susceptibility anisotropy
due to the combination of nematic/orbital order and spin-orbit coupling
is much weaker than the one caused by the anisotropy due to long range
magnetic order. Thus, the combination of susceptibility and resistivity
anisotropy can be used to disentangle these two phenomena.

The most natural way to account for the anisotropy of the magnetic
susceptibility in the magnetically ordered state is to include spin-orbit
coupling in the effective low-energy model of the iron-based superconductors.
Indeed, it is responsible for the observed magnetic anisotropy of
the striped antiferromagnetic state, namely, for the alignment of
the magnetic moments parallel to the AF wave-vector $\mathbf{Q}_{1}$
at the transition temperature \cite{Christensen2015}. We describe
(details can be found in supplemental material \cite{[{See Supplemental Material at URL will be inserted by publisher for calculation details}]Supplemental})
the itinerant electron system of the parent iron-based superconductors
by a multi-orbital Hubbard Hamiltonian, which consists of the non-interacting
hopping Hamiltonian within the $3d$-orbital manifold, $H_{0}$, and
Hubbard-Hund interaction, $H_{\mathrm{int}}$. We specify the hopping
parameters $t_{ij}^{\mu\nu}$ according to the band-structure parametrization
obtained by Ikeda \cite{Ikeda2010} for a five orbital model or three-orbital
model by Daghofer \cite{Daghofer2010}. Besides the band dispersions,
the non-interacting Hamiltonian must also contain the spin orbit coupling
term $\lambda{\bf S}\cdot{\bf L}$ with ${\bf S}$ and ${\bf L}$
denoting the spin and orbital angular momentum operator, respectively.
Note that this atomic-like term preserves the Kramers degeneracy of
each state. We project this term from the L = 2 spherical harmonic
basis to the orbital basis using the standard procedure of Ref. \cite{Christensen2015}.
In order to simulate the breaking of the $C_{4}$ symmetry above $T_{N}$
in the experiment, we also introduced a uniform energy splitting of
the $d_{xz}$ and $d_{yz}$ orbitals \cite{Fernandes2014PRB}, 
\begin{equation}
H_{oo}=\Delta_{oo}\sum_{\mathbf{k\mathbf{\sigma}}}\bigl(c_{xz\mathbf{k\sigma}}^{\dagger}c_{xz\mathbf{k}\sigma}-c_{yz\mathbf{k\sigma}}^{\dagger}c_{yz\mathbf{k}\sigma}\bigr),\label{eq:3}
\end{equation}
where $\Delta_{oo}$=-25 meV was used so that $d_{yz}$ shifts upwards.
Note that such a term appears in the striped AF state automatically
as a result of the magnetic ordering breaking the $C_{4}$ symmetry
of the lattice.

\begin{figure}
\begin{centering}
\includegraphics[scale=0.50]{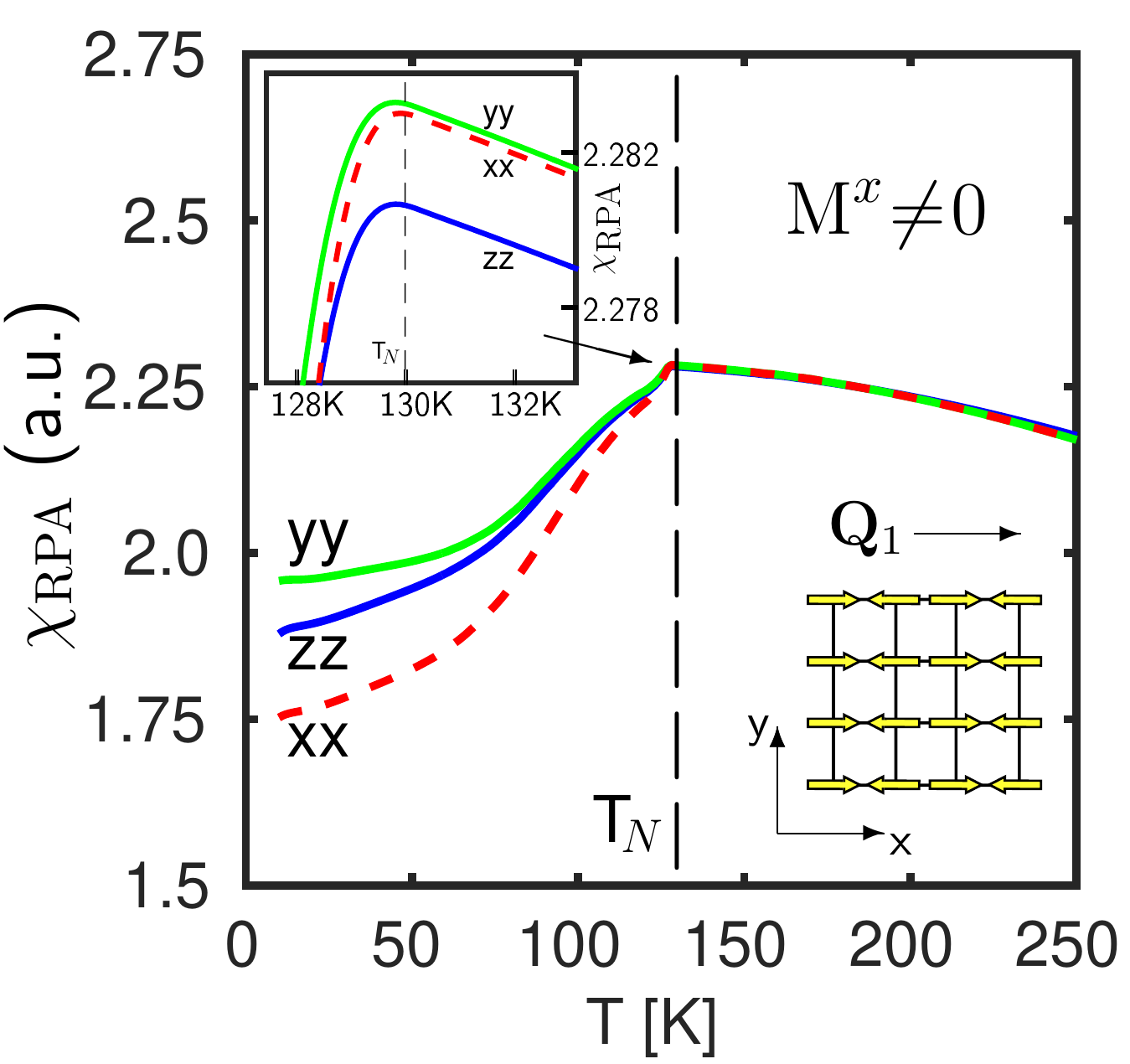}
\par\end{centering}
\caption{Magnetic susceptibility calculated in the stripe AF phase using an
itinerant multi-orbital model(coordinate basis is transformed as $a\rightarrow x,b\rightarrow y,c\rightarrow z$
in comparison with experimental data). The magnetic moments are arranged
parallel to the AF wave vector $\mathbf{Q}_{1}$ so that $M^{x}\protect\neq0,M^{y,z}=0$
resulting in $\chi{}^{yy}>\chi{}^{xx}$ in agreement with experiment
. The inset shows an enlarged view near the transition where an extremely
weak splitting ($\ll$1\%) between $\chi{}^{xx}$ and $\chi{}^{yy}$
occurs in the paramagnetic state due to the finite orbital ordering
($\Delta_{oo}$=-25 meV ).\label{fig:4}}
\end{figure}

The results of our susceptibility calculations (see supplemental material
for details \cite{[{See Supplemental Material at URL will be inserted by publisher for calculation details}]Supplemental}) are shown in Fig. \ref{fig:4}.
To compare with experimental data, we assign $a\rightarrow x,b\rightarrow y,c\rightarrow z$.
As expected, the sign of in-plane susceptibility anisotropy strongly
depends on the orientation of the magnetic moments. Alignment of the
magnetic moments along $\mathbf{Q}{}_{1}$ driven by spin-orbit coupling
produces the anisotropy observed in our magnetization experiments,
i.e. $\chi^{yy}>\chi^{xx}$. We note that this is also the same anisotropy
expected in a purely localized magnetic model, i.e. the susceptibility
is larger for fields perpendicular to the moments. Apart from spin-orbit
coupling, the calculation shows that the Umklapp susceptibility dominated
by intra-orbital ($yz$,$yz$) contributions is responsible for the
observed pronounced anisotropy. The inset in Fig. \ref{fig:4} shows
that the anisotropy induced by finite orbital ordering in the paramagnetic
state is extremely weak $\eta=\chi^{yy}/\chi^{xx}-1\ll$1\%. Orbital
ordering therefore can not be responsible for the non-negligible anisotropy
slightly above $T_{N}$ observed in our experimental data. The small
effect is however in agreement with the comparatively small orbitally
induced susceptibility anisotropy in the wide region between 150K
and 200K, see Fig. \ref{fig:3}.

In summary, we have determined the in-plane resistivity and susceptibility
anisotropies of BaFe\textsubscript{2}As\textsubscript{2} using a
new and simple method, which applies a large uniaxial strain. 
Interestingly, in spite of the strain-induced 'smearing' of the structural,
or nematic transition, the resistivity anisotropy and its corresponding
nematic susceptibility show the same behavior as those measured in
zero-strain limit, suggesting that the resistivity anisotropy is more
directly related to the magnetic transition than to the nematic fluctuations.
The observed susceptibility anisotropy in the magnetically ordered
phase qualitatively agrees well with calculations using an effective
low-energy itinerant model including spin-orbit coupling, in which
the sizable splitting is dominated by intra-orbital $(yz,yz)$ Umklapp
processes. Striking is the different behavior of the resistivity and
susceptibility anisotropies in the paramagnetic uniaxially strained
state. In particular, whereas the resistivity anisotropy exhibits
a Curie-Weiss divergence extending to temperatures much larger than
$T_{N}$, the susceptibility anisotropy develops only about 10 K above
$T_{N}$. Our calculations show that orbital order produces a negligible
susceptibility anisotropy above $T_{N}$ and serve to disentangle
anisotropies due to orbital and nematic order from those of the magnetically
ordered state. 

We thank Rafael Fernandes and Igor Mazin for valuable discussions.

%

  \end{bibunit}

\clearpage{}

\section{Supplemental Material: Dichotomy between in-plane magnetic susceptibility
and resistivity anisotropies in extremely strained BaFe\protect\textsubscript{2}As\protect\textsubscript{2} }
\begin{bibunit}[apsrev4-1]
\renewcommand{\thefigure}{S\arabic{figure}} 
\setcounter{figure}{0} 
\renewcommand{\theHfigure}{Supplement.\thefigure}

The itinerant electron system of the parent iron-based superconductors
can be described by a multi-orbital Hubbard Hamiltonian, which consists
of the non-interacting hopping Hamiltonian within the $3d$-orbital
manifold, $H_{0}$, and the Hubbard-Hund interaction, $H_{\mathrm{int}}$,
\begin{equation}
H=H_{0}+H_{\mathrm{int}},\label{eq:hamiltonian}
\end{equation}
with 
\begin{equation}
H_{0}=\sum_{\sigma}\sum_{i,j}\sum_{\mu,\nu}c_{i\mu\sigma}^{\dagger}\left(t_{ij}^{\mu\nu}-\mu_{0}\delta_{ij}\delta_{\mu\nu}\right)c_{j\nu\sigma},\label{eq:hopping}
\end{equation}
and 
\begin{eqnarray}
H_{\mathrm{int}} & = & U\sum_{i,\mu}n_{i\mu\uparrow}n_{i\mu\downarrow}+U^{\prime}\sum_{i,\mu<\nu,\sigma}n_{i\mu\sigma}n_{i\nu\bar{\sigma}}+\label{eq:interaction}\\
 &  & (U^{\prime}-J)\sum_{i,\mu<\nu,\sigma}n_{i\mu\sigma}n_{i\nu\sigma}+\nonumber \\
 &  & J\sum_{i,\mu<\nu,\sigma}c_{i\mu\sigma}^{\dagger}c_{i\nu\bar{\sigma}}^{\dagger}c_{i\mu\bar{\sigma}}c_{i\nu\sigma}+\nonumber \\
 &  & J^{\prime}\sum_{i,\mu<\nu,\sigma}c_{i\mu\sigma}^{\dagger}c_{i\mu\bar{\sigma}}^{\dagger}c_{i\nu\bar{\sigma}}c_{i\nu\sigma}.\nonumber 
\end{eqnarray}
The indices $\mu,\nu\in\{d_{xz},d_{yz},d_{x^{2}-y^{2}},d_{xy},d_{3z^{2}-r^{2}}\}$
specify the $3d$-Fe orbitals and $i,j$ run over the sites of the
square lattice. The doping is fixed by the chemical potential $\mu_{0}$.
The interactions are parametrized by an intra-orbital on-site Hubbard-$U$,
an inter-orbital coupling $U^{\prime}$, Hund's coupling $J$ and
pair hopping $J^{\prime}$. We employ $U^{\prime}=U-2J$, $J=J^{\prime}$
and set $J=U/4$. The fermionic operators $c_{i\mu\sigma}^{\dagger}$,
$c_{i\mu\sigma}$ are the creation and annihilation operators, respectively.
We specify the hopping parameters $t_{ij}^{\mu\nu}$ according to
the band structure obtained by Ikeda \textit{et al.} \cite{Ikeda2010s}
or Daghofer et al. \cite{Daghofer2010s} for the five or three orbital
models, respectively.

\begin{figure}
\begin{centering}
\includegraphics[scale=0.45]{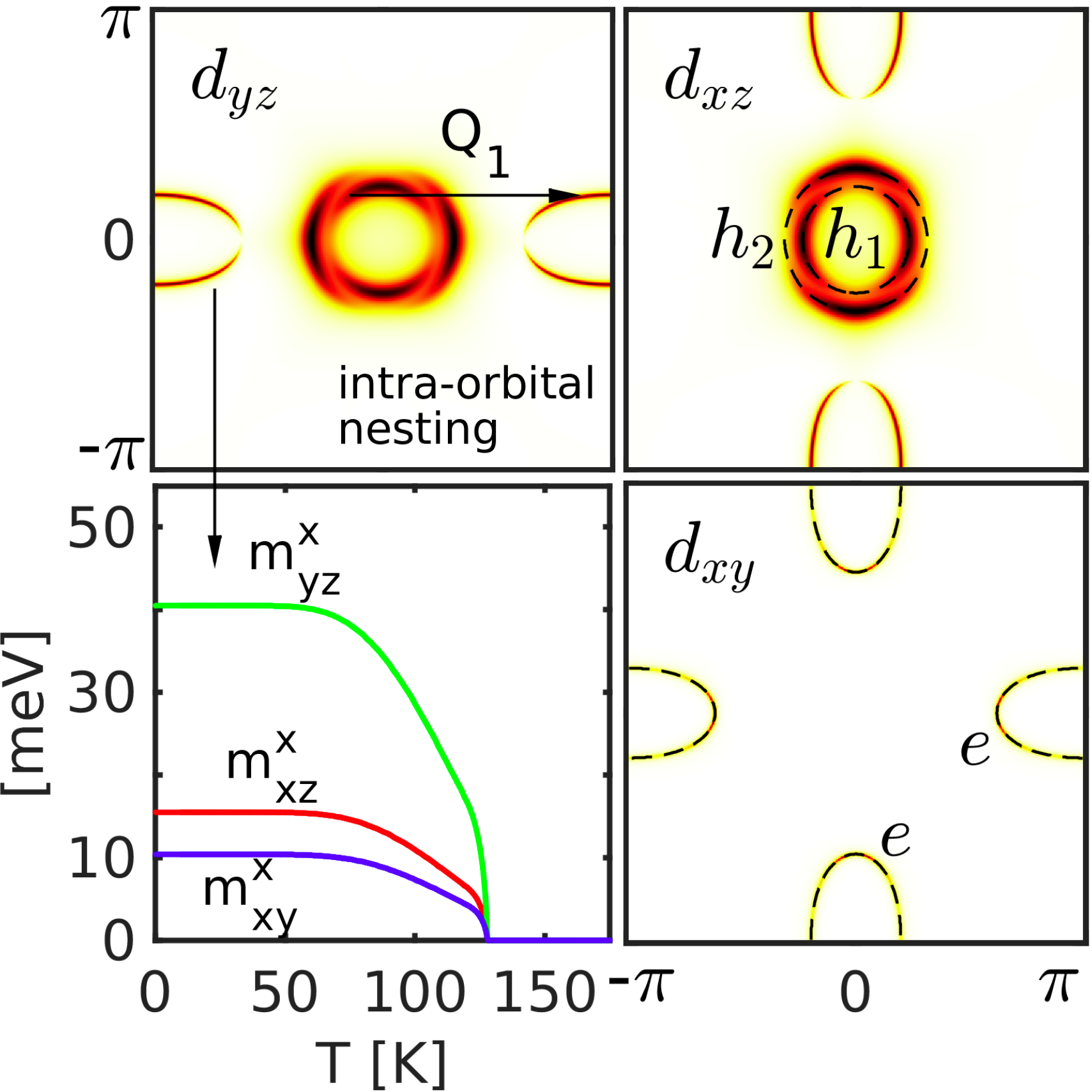}
\par\end{centering}
\caption{Typical distribution of the orbital content at the Fermi surface for
the iron-based compounds \cite{Ikeda2010s,Daghofer2010s}. For the wave-vector
$\mathbf{Q}_{1}$, the intra-orbital nesting of the $d_{yz}$-orbitals
dominates over the other orbitals contributions. As a direct consequence,
the SDW gap (bottom left corner) and the magnetic Umklapp susceptibility
will acquire the largest contribution due the $d_{yz}$-orbital. The
magnetization of the other two orbitals, present at the Fermi level,
is mainly induced by the Hund coupling $J$. Here, $m_{l}^{i}$ corresponds
to the magnetization of the $l=yz,xz,xy$ orbital along the spin orientation
$i=x,y,z$ .\label{fig:FS}}
\end{figure}

The approximate nesting between hole pockets around $\Gamma$ and
$M$ and electron pockets around $X$ and $Y$ promotes strong fluctuations
in the particle-hole channel at wave vectors ${\bf Q}_{1}=(\pi,0)$
and ${\bf Q}_{2}=(0,\pi)$. The electronic states at the Fermi level
are dominated by the $d_{xz}$, $d_{yz}$ and $d_{xy}$ orbitals,
as shown in Fig. \ref{fig:FS}. While the hole-pockets centered around
$\Gamma$ are formed by $d_{xz}$ and $d_{yz}$ orbitals, the hole
pocket at $M$ is mainly of $d_{xy}$ character. The electron pockets
at $X$ and $Y$ feature a mixed orbital character, where the inner
part facing towards the BZ center is $d_{xy}$ dominated, and the
outer parts are $d_{yz}$ and $d_{xz}$ dominated around $(\pi,0)$
and $(0,\pi)$, respectively. This allows for the further reduction
of the 5-orbital model to the three-orbital model only \cite{Daghofer2010s}.

This directionality of the SDW order parameter, on top of the breaking
of rotational symmetry in spin space, also breaks the $C_{4}$ symmetry
of the five orbital model down to a $C_{2}$ symmetry. The two different,
orbitally resolved SDW order parameters read as ${\bf M}_{1}^{\mu\nu}=\frac{1}{\mathcal{N}}\sum_{{\bf k},\sigma,\sigma^{\prime}}\langle c_{{\bf k}+{\bf Q}_{1}\mu\sigma}^{\dagger}{\boldsymbol{\sigma}}_{\sigma\sigma^{\prime}}c_{{\bf k}\nu\sigma^{\prime}}\rangle$,
and ${\bf M}_{2}^{\mu\nu}=\frac{1}{\mathcal{N}}\sum_{{\bf k},\sigma,\sigma^{\prime}}\langle c_{{\bf k}+{\bf Q}_{2}\mu\sigma}^{\dagger}{\boldsymbol{\sigma}}_{\sigma\sigma^{\prime}}c_{{\bf k}\nu\sigma^{\prime}}\rangle,$
with $\mathcal{N}$ the number of unit cells and ${\boldsymbol{\sigma}}$
the vector of Pauli matrices. Here, the fermionic creation and annihilation
operators for orbital Bloch states with wave vector ${\bf k}$ are
defined as $c_{{\bf k}\mu\sigma}^{\dagger}=\frac{1}{\sqrt{\mathcal{N}}}\sum_{i}\mathrm{e}^{-\mathrm{i}{\bf k}\cdot{\bf r}_{i}}c_{i\mu\sigma}^{\dagger}$,
$c_{{\bf k}\mu\sigma}=\frac{1}{\sqrt{\mathcal{N}}}\sum_{i}\mathrm{e}^{\mathrm{i}{\bf k}\cdot{\bf r}_{i}}c_{i\mu\sigma}$.
Taking the orbital trace yields the magnetic moments ${\bf M}_{1}$,
${\bf M}_{2}$ of the two SDW configurations.

Besides the band dispersions, the non-interacting Hamiltonian must
also contain the SOC term $\lambda{\bf S}\cdot{\bf L}$, with ${\bf S}$
denoting the spin angular momentum operator and ${\bf L}$, the orbital
angular momentum operator, projected from the $L=2$ cubic harmonic
basis to the orbital basis \cite{Christensen2015s}. Furthermore, in
the following we assume that the system possesses a striped AF order
with ${\bf Q}_{1}$ ordering wave-vector and the magnetic moment is
pointing along the ordering momentum, i.e. $x$-direction. Such an
order appears to be the ground state for zero doping in several studies
of the typical models of the iron-based superconductors \cite{Gastiasoro2014s,Christensen2015s}.
To understand the origin of anisotropy in the uniform susceptibility
in the magnetic state, we note that the magnetic inter-orbital components
of the mean-field magnetizations for the $C_{2}$ phase were found
to be negligible compared to the intra-orbital terms \cite{Gastiasoro2014s}.
Furthermore, as mentioned above there are three orbitals contributing
to the Fermi surfaces, however, only one of them has a significant
portion of the intra-orbital nesting.

As a result the magnetization for the striped antiferromagnetic state
with ${\bf Q}_{1}$ wave vector has largest contribution that arises
from the $yz$-orbital, as shown in Fig. \ref{fig:FS}.

Next we compute the components of the magnetic susceptibility $\chi_{0}^{xx/yy/zz}$
in the multi-orbital case, 
\begin{multline}
\chi_{0}^{uu}(\mathbf{q},\omega)=-\sum_{\mathbf{k}ij}\eta^{uu}(i,\mathbf{k};j,\mathbf{k+q})\times\\
\frac{f(E_{j}(\mathbf{k+q}))-f(E_{i}(\mathbf{k}))}{E_{j}(\mathbf{k+q})-E_{i}(\mathbf{k})+\omega+i0^{+}},\quad u=x,y,z,\label{eq:chi}
\end{multline}
which includes the tensor 
\begin{multline}
[\eta^{uv}(i,\mathbf{k};j,\mathbf{k+q})]_{q\beta,p\alpha}^{s\gamma,t\delta}=\sigma_{\alpha\beta}^{u}\sigma_{\gamma\delta}^{v}\times\\
a_{q\beta}(i,\mathbf{k})a_{s\gamma}^{*}(i,\mathbf{k})a_{t\delta}(j,\mathbf{k+q})a_{p\alpha}^{*}(j,\mathbf{k+q}),\label{eq:dressing}
\end{multline}
expressed with the help of the Pauli matrices $\sigma^{u}$ and the
the elements of the unitary transformations from the band to the orbital
basis, $a$. The physical susceptibility is then obtained by taking
the trace over $p=q$ and $s=t$ orbitals.

It is straightforward to show that the anisotropy of the susceptibility
enters through the orbital-dressing factors (\ref{eq:dressing}) and
presence of the Umklapp terms in the antiferromagnetic translational-symmetry
broken state. We find for the bare susceptibility: 

\begin{align}
\begin{split}\chi_{0}^{xx}(\mathbf{Q,\omega}) & \!=\!\diagram{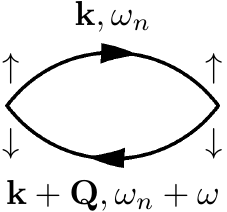}+\diagram{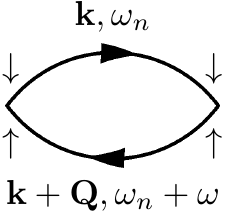}+\diagram{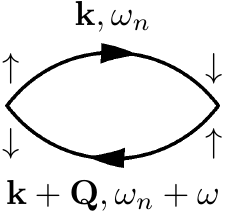}+\diagram{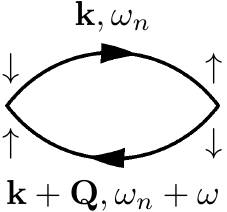}\\
\chi_{0}^{yy}(\mathbf{Q,\omega}) & \!=\!\diagram{bubbleuudd}+\diagram{bubbledduu}-\diagram{bubbleudud}-\diagram{bubbledudu}\\
\chi_{0}^{zz}(\mathbf{Q,\omega}) & \!=\!\diagram{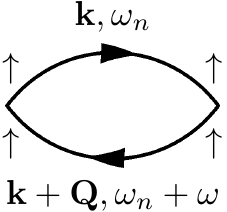}+\diagram{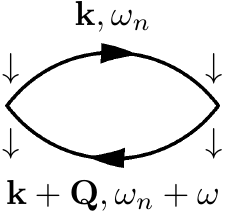}-\diagram{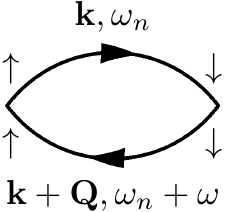}-\diagram{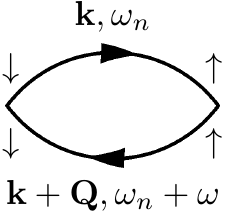}
\end{split}
\label{eq:bubbles}
\end{align}
Note that in the paramagnetic state and for vanishing spin-orbit coupling,
the first two bubbles of each component are equal while the last two
Umklapp terms vanish, ensuring the overall $O(3)$ symmetry of the
system. If spin-orbit coupling is finite, we find $\chi^{yy}>\chi^{xx,zz}$
at $\mathbf{Q}_{1}=(\pi,0)$ so that an alignment of the magnetic
moments parallel to $\mathbf{Q}_{1}$ is favored (M$^{x}\neq0$, M$^{y,z}=0$).

Although spin-orbit coupling is large enough to lower the symmetry
by favoring M$^{x}$ over M$^{y}$ for the AF wave-vector $\mathbf{Q}_{1}$,
it is not large enough to account for the size of the in-plane anisotropy
observed below the AF transition temperature. In the following we
show that the in-plane splitting of the uniform susceptibility is
caused by the intra-orbital Umklapp terms of the $yz$ orbital ($xz$
respectively for $\mathbf{Q}_{2}$). In particular, looking at the
Eq.(\ref{eq:bubbles}), one sees that the difference of the two in-plane
components originates from the third and fourth bubble diagrams due
to the different sign for $\chi^{xx}$ and $\chi^{yy}$, 
\begin{align}
G_{yz,\sigma;yz,\bar{\sigma}}(i,\mathbf{k})G_{yz,\sigma;yz,\bar{\sigma}}(j,\mathbf{k+q+Q}_{1}).\label{eq:umklapp}
\end{align}
Evaluating the sums we find for the splitting 
\begin{widetext}
\begin{align}
[\chi_{0}^{xx}-\chi_{0}^{yy}]_{(\mathbf{q}=0)} & =4\Bigl[(m_{yz}^{x})^{2}-(m_{yz}^{y})^{2}\Bigr]\sum_{i\mathbf{k}}|a_{yz\uparrow}(h_{i},\mathbf{k})|^{4}|a_{yz{\downarrow}}(e,\mathbf{k+Q}_{1})|^{4}\;\frac{f(E_{i}(\mathbf{k}))-f(E_{e}(\mathbf{k+Q}_{1}))}{\bigl[E_{i}(\mathbf{k})-E_{e}(\mathbf{k+Q}_{1})\bigr]^{3}}\label{eq:anisotropy}
\end{align}
\end{widetext}

where we have set $\omega=0$ and denote $m_{l}^{i}$ to be the magnetization
of the $l=yz,xz,xy$ orbital along the spin orientation $i=x,y,z$.
As one clearly sees in the stripe AF phase with ordering momentum
$\mathbf{Q}_{1}$ with spins aligned either parallel or antiparallel
to the %
\mbox{%
x-direction%
} %
\mbox{%
$(\text{M}^{x}\neq0,m_{yz}^{x}\neq0)$%
}, both transverse components of the susceptibility split and one has
$\chi_{0}^{yy}(\mathbf{q}=0,\omega=0)>\chi_{0}^{xx}(\mathbf{q}=0,\omega=0)$.
Furthermore, the sign of the anisotropy is reversed %
\mbox{%
$(\text{M}^{y}\neq0,m_{yz}^{y}\neq0$%
}) if the moments would be pointing out perpendicular to the ordering
wave vector. In simple terms the largest AF gap in the spin subspace
reduces the corresponding component of the uniform susceptibility.

Our analytical results for the $yz$ orbitals are fully confirmed
by the full numerical calculations using the realistic tight-binding
models \cite{Ikeda2010s,Daghofer2010s}. In particular, in Fig. 4 of
the main text we show the calculated uniform susceptibility splitting
calculated within random phase approximation. In addition, the numerical
study confirms that the in-plane anisotropy is determined by the Umklapp
susceptibility involving $yz$ and that the sign of the anisotropy
depends on the orientation of the magnetic moments.

Finally we note by passing that the origin of the magnetic anisotropy
in the uniform susceptibility cannot be due to the simple ferro-obital
ordering ($n_{xz}-n_{yz}\neq0$), introduced by the structural (nematic)
transition at $T_{S}>T_{N}$ although it breaks the anisotropy between
the $x$ and the $y$ component of the spin susceptibility. One finds
in this case that the splitting will be proportional to $\Delta_{\text{oo}}\lambda^{2}$
and thus its sign is reversed as compared to the effect of the magnetic
ordering. 
%

  \end{bibunit}
\end{document}